\documentclass{pnastwo}
\usepackage[dvips]{graphicx}
\usepackage{pnastwoF}
\usepackage{amssymb,amsfonts,amsmath}
\usepackage{graphicx}
\usepackage{dcolumn}
\usepackage{epstopdf}
\usepackage{bm}
\usepackage{graphicx}
\usepackage{subfigure}
\usepackage{color}
\usepackage{latexsym}
\usepackage{amsfonts}
\usepackage{amssymb,txfonts, pxfonts}
\usepackage{amssymb,amsfonts,amsmath}

                \def \e{\epsilon}

  \def \t{\bar{t}}   % for writing partialderivatives

  \def \>{\rangle} \def \<{\langle}
 \def\[{\left[} \def\]{\right]} 
   
     \def\p{\tilde{p}}

\def\p{\partial} \def\et{{\it et al}}

\def\ecoli{{\it E. coli }}
\def\max{\mathrm{max}}

 \def\lam{{\lambda}}

  \def\be{\begin{equation}} \def\ee{\end{equation}}
\newcommand \bea {\begin{eqnarray} } \newcommand \eea {\end{eqnarray}}

\newcommand{\mon}{\begin{displaymath}}
\newcommand{\moff}{\end{displaymath}}

\newcommand{\od}[2]{\frac{d {#1}}{d {#2}}}
\newcommand{\eon}{\begin{equation}}
\newcommand{\eoff}{\end{equation}}
\newcommand{\eaon}{\begin{eqnarray}}
\newcommand{\eaoff}{\end{eqnarray}}

\newcommand{\eq}[1]{Eq.~(\ref{#1})}

\newcommand{\fig}[1]{Fig.~\ref{#1}}
\newcommand{\ue}{U_b}
\newcommand{\ud}{U_d}

\renewcommand{\ss}{s}
\def \D{\ss}
\def \hs{\hat{s}}
\newcommand{\avgfit}{\bar{\fit}}
\newcommand{\fit}{{\omega}}
\newcommand{\nz}{n_{k_\star}}
\url{www.pnas.org/cgi/doi/....}
\copyrightyear{2011}
\issuedate{Issue Date}
\volume{Volume}
\issuenumber{Issue Number}
\begin{document}

\title{%Rare beneficial mutations halt Muller's ratchet
Rare beneficial mutations can halt Muller's ratchet
%Rare beneficial mutations can prevent mutational meltdown
}

\author{Sidhartha Goyal\affil{1}{Kavli Institute for Theoretical Physics,
University of California, Santa Barbara, CA}, Daniel J.~Balick\affil{2}{Department
of Physics, University of California, Santa Barbara, CA}, Elizabeth R.
Jerison\affil{3}{Departments of Organismic and Evolutionary Biology and of Physics, and FAS Center for Systems Biology, Harvard University, Cambridge MA}, Richard A.~Neher\affil{4}{Max Planck
Institute for Developmental Biology, T\"ubingen, Germany}, Boris I.~Shraiman\affil{1}{Kavli Institute for Theoretical Physics, University of California, Santa Barbara, CA}\!\!\affil{2}{Department
of Physics, University of California, Santa Barbara, CA}, and
Michael M.~Desai\affil{3}{Departments of Organismic and Evolutionary Biology and of Physics, and FAS Center for Systems Biology, Harvard University, Cambridge MA}}

\contributor{Submitted to Proceedings of the National Academy of Sciences of the United States of America}

\maketitle

\begin{article}
\begin{abstract}
The vast majority of mutations are deleterious, and are eliminated by purifying selection. Yet in finite asexual populations, purifying selection cannot completely prevent the accumulation of deleterious mutations due to Muller's ratchet: once lost by stochastic drift, the most-fit class of genotypes is lost forever. If deleterious mutations are weakly selected, Muller's ratchet turns into a mutational ``meltdown'' leading to a rapid degradation of population fitness. Evidently, the long term stability of an asexual population requires an influx of beneficial mutations that continuously compensate for the accumulation of the weakly deleterious ones. Here we propose that the stable evolutionary state of a population in a static environment is a dynamic mutation-selection balance, where accumulation of deleterious mutations is on average offset by the influx of beneficial mutations. We argue that this state exists for any population size $N$ and mutation rate $U$.  Assuming that beneficial and deleterious mutations have the same fitness effect $\ss$, we calculate the fraction of beneficial mutations, $\e$, that maintains the balanced state. We find that a surprisingly low $\e$ suffices to maintain stability, even in small populations in the face of high mutation rates and weak selection.  This may explain the maintenance of mitochondria and other asexual genomes, and has implications for the expected statistics of genetic diversity in these populations.
\end{abstract}

\keywords{asexual evolution | Muller's ratchet | adaptation}

\dropcap{P}urifying selection maintains well-adapted genotypes in the face of deleterious mutations \cite{Haigh78}.  Yet in asexual populations, random genetic drift in the most-fit class of individuals will occasionally lead to its irreversible extinction, a process known as Muller's ratchet \cite{Muller64, Felsenstein74}.  The repetitive action of the ratchet leads to the accumulation of deleterious mutations, despite the action of purifying selection.  This ratchet effect has been extensively analyzed \cite{Gordo00, Gordo00b, Stephan02, Charlesworth97, Gessler95} and has been observed in experiments \cite{Chao90, Duarte92, Andersson96, Zeyl01} and in nature \cite{Rice94, Lynch96, Howe08}. In small populations when deleterious mutation rates are high or selection pressures are weak, the ratchet can proceed quickly, causing rapid degradation of asexual genomes and a collapse in population size: the so-called "mutational meltdown" \cite{Lynch93, Gabriel93, Lynch95}.  Hence, Muller's ratchet has been described as a central problem for the maintenance of asexual populations such as mitochondria \cite{Loewe06}. Avoiding this mutational catastrophe is thought to be a major benefit of sex and recombination (see \cite{Barton98, DeVisser07} for reviews).

\begin{figure}[tp]
\begin{center}
\includegraphics[width=0.7\columnwidth]{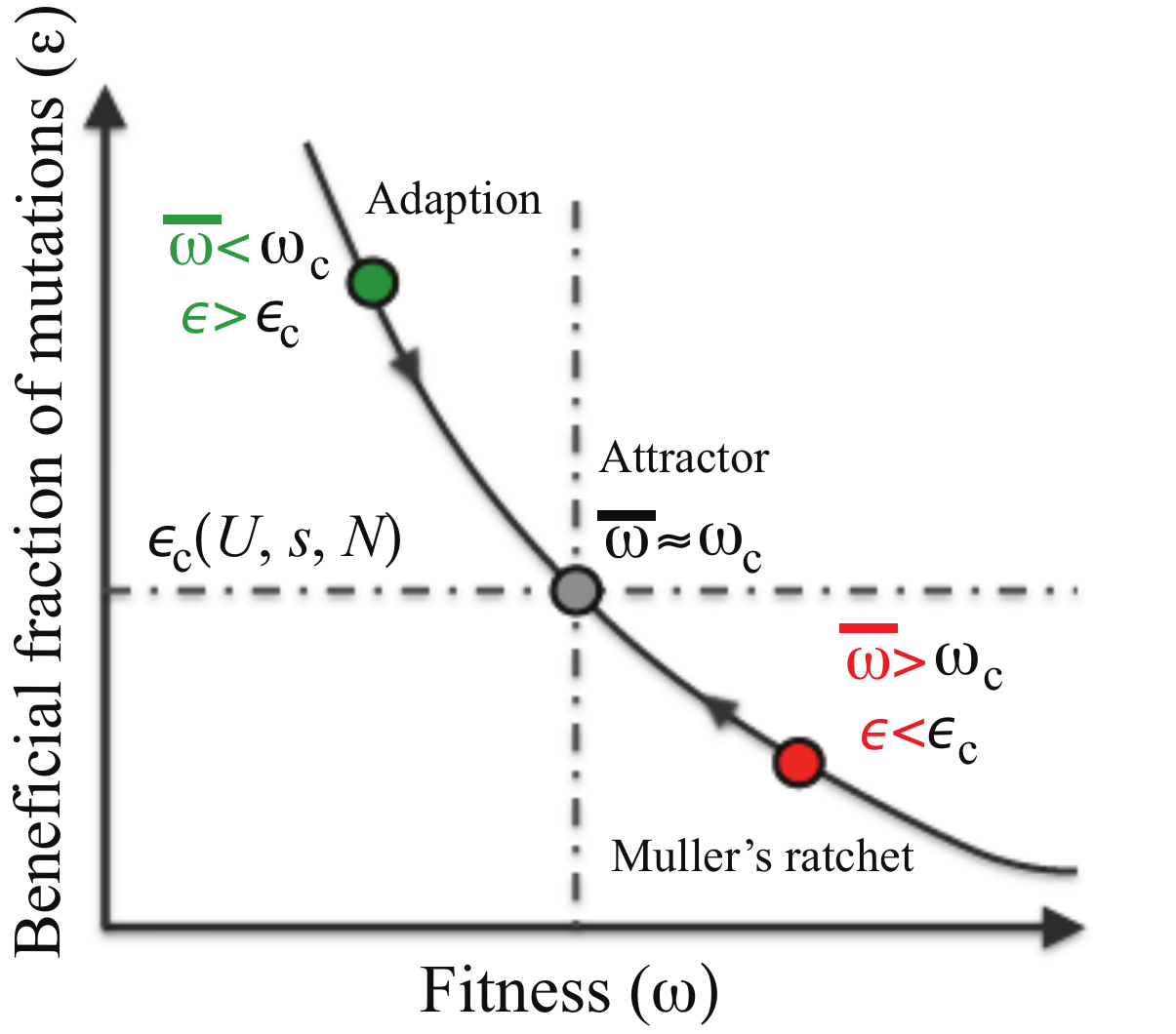}
\caption{\label{fig:fitlandscape} The relationship between the fraction of mutations that are beneficial, $\e$, and the absolute fitness of the population, $\fit$.  A poorly adapted population (green) with $\avgfit < \fit_c$ and $\e > \e_c$, will adapt towards the dynamic equilibrium "Attractor" state (gray) with higher fitness and lower $\e$.  Conversely, an ``over-adapted'' population (red) with $\avgfit > \fit_c$ and $\e < \e_c$, will decline in fitness towards the dynamic equilibrium state due to Muller's ratchet.  }
\end{center}
\end{figure}

In the absence of recombination \cite{Bell88} or epistasis \cite{Kondrashov94}, the only forces that can check the accumulation of deleterious mutations are back and compensatory mutations.  These mutations are typically assumed to be rare and are either neglected in models of the ratchet, or assumed to only slightly slow its rate \cite{Haigh78, Kondrashov95, Gordo04}.  However, extensive experimental work \cite{Barrick10, Schoustra09, Silander07, Escarmis99, Poon05a, Poon05b, Lynch03} and theoretical expectations  \cite{Fisher30, Orr03, Gillespie84, Callahan11} suggest that the number of available back and compensatory mutations increases as a population accumulates deleterious mutations and declines in fitness.  The increase in beneficial mutation rates with declining fitness will eventually stop the progress of Muller's ratchet, as illustrated in \fig{fig:fitlandscape}.  Furthermore, since back and compensatory mutations are selectively favored, they can balance deleterious mutations even while they are relatively rare, thereby maintaining a well-adapted population.

Despite the potential relevance of compensatory mutations, only a few studies have quantitatively analyzed the simultaneous accumulation of both beneficial and deleterious mutations.  These include simulations by Antezana and Hudson \cite{Antezana97} and Wagner and Gabriel \cite{Wagner90}, experimental work by Silander \et. \cite{Silander07}, and both simulations and analytical work by Rouzine \et.~\cite{Rouzine03, Rouzine08}. In each of these analyses, simulations or experiments showed that the mean population fitness settles at a steady-state value $\fit_c$ --- the point where beneficial mutations arise often enough to compensate for newly-fixed deleterious mutations.

While the absolute fitness, $\fit_c$, of the population in this dynamic equilibrium state is model dependent, the existence of the stable steady state is generic.  Intuitively, suppose a fraction $\e$ of mutations are beneficial, and the critical fraction needed to maintain stability is $\e_c$. In a poorly adapted population, a higher fraction of mutations will be beneficial, $\e > \e_c$. This excess of beneficial mutations will push the population towards higher fitness. At the same time, adaptation will deplete the available pool of beneficial mutations, until $\e_c$ is reached, as shown in \fig{fig:fitlandscape}.  Conversely, in an ``over-adapted'' population,  we expect $\e < \e_c$, so that deleterious mutations dominate, reducing population fitness until $\e_c$ is recovered. Thus we expect the dynamic mutation-selection balance point to be a stable evolutionary ``attractor.'' In contrast to $\fit_c$, the critical $\e_c$ provides a general measure of how well adapted a population is, independent of the model of compensatory mutations.

Below, we show analytically how $\e_c$ depends on population size, mutation rate, and strength of selection and compare our predictions  to numerical simulations.  We show that even for high mutation rate and small population size, the stable state occurs at high fitness, where most mutations are deleterious.  Muller's ratchet notwithstanding, selection enables rare beneficial mutations to compensate for more frequent deleterious mutations, and to maintain a well adapted population.

\section{Model}

\begin{figure}[tp]
\begin{center}
\includegraphics[width=1\columnwidth]{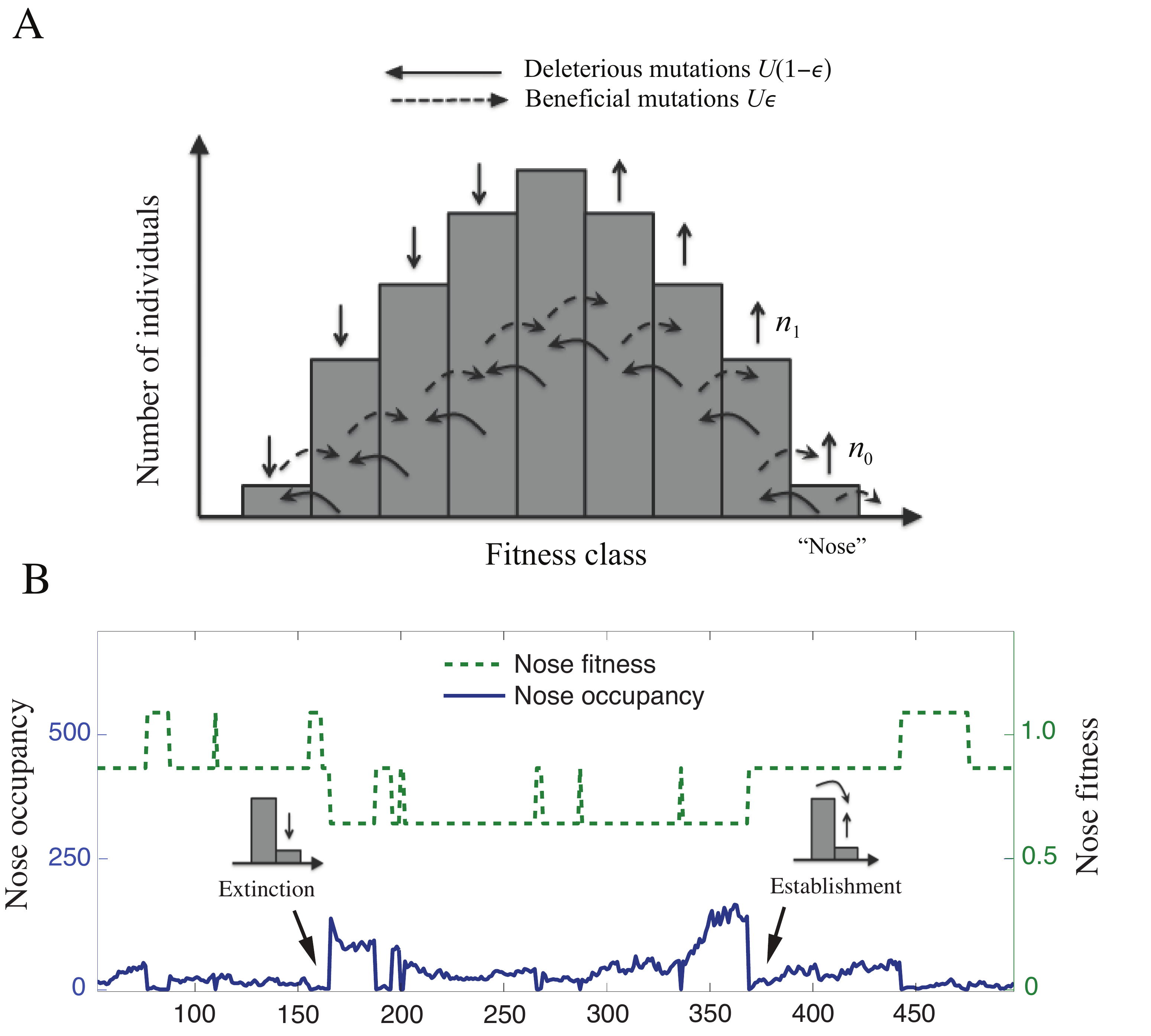} 	
\caption{\textbf{(A)} Schematic illustration of the fitness distribution within a population.  We refer to the most-fit class as the ``nose'' of the distribution. \textbf{(B)} A typical realization of the stochastic dynamics at the nose. In the event of extinction of the nose, $n_1$ becomes the new nose and relative nose fitness decreases by $\ss$. Conversely, when a more-fit nose is established the relative nose fitness increases by $\ss$. \label{fig:model}}
\end{center}
\end{figure}

Our analysis is based on the standard discrete generation Wright-Fisher model with population size fixed at $N$.  We assume that beneficial and deleterious mutations increase or decrease fitness by $\ss$ (where $\ss > 0$ by convention). Each individual has some number, $k$, deleterious mutations relative to the perfectly adapted state.  We define $\bar{k}$ to be the mean number of deleterious mutations per individual, so that an individual with $k$ deleterious mutations has fitness~$\D (\bar{k} - k)$ relative to the mean, as shown in Fig.~\ref{fig:model}.  The population can be characterized by the number of individuals $n_{k}$ in each of these fitness classes.  We define $U$ to be the total mutation rate, and $\e$ to be the fraction of mutations that are beneficial, so that the beneficial mutation rate is $\ue = U \e$, and the deleterious mutation rate is $\ud = U (1-\e)$.

Although mutations and genetic drift are stochastic processes, in the bulk of the distribution --- where fitness classes contain many individuals --- the population dynamics are well-captured by a deterministic approximation.  On average, the number of individuals in fitness class $k$ evolves as: \be \frac{dn_{k}}{dt} = \underbrace{\D(\bar{k} - k)n_{k}}_{I} - \underbrace{U n_{k}}_{II} + \underbrace{U_dn_{k-1}}_{III} + \underbrace{ U_b n_{k+1}}_{IV} \label{det} . \ee The terms on the right hand side of this equation describe the different processes acting on fitness class $k$:  (I) the effect of selection relative to the mean fitness, (II) mutational load, (III) deleterious mutations from more-fit individuals, and (IV) beneficial mutations from less-fit individuals.  We assume that the selective effect of a single mutation is small, $\ss \ll 1$.

In general, we expect $\e$ to increase with $k$, since less-fit genotypes have a higher frequency of beneficial back and compensatory mutations.  However, at any given time there will only be a relatively narrow range of $k$ present in the population.  Thus in \eq{det} we neglect the $k$-dependence of $\e$; we have tested this approximation against a simulation incorporating the $k$-dependence of $\e$ (see SI).  We use this approximation to compute the $\e_c$ and corresponding distribution $n_{k}$ at which the population stays on average at a constant fitness. As we have argued above, the stability of this equilibrium state is ensured by the $k$-dependence of $\e$ on larger scales.

\section{Results}
In steady state, the fitness distribution $n_{k}$  stays constant on average: $\od{n_{k}}{t} = 0$ for all $k$. Solving \eq{det} for the steady state in the case of $\e = 0$ leads to the familiar mutation-selection balance \cite{Haigh78}, with a Poisson distribution of fitness: $n_{k} = N e^{-\lambda}{\lambda^{k} \over k!}$, where $\lambda =U/\D$ and $k=0$ corresponds to the most-fit class.

This deterministic approximation corresponds to an infinite population where Muller's ratchet (i.e. the stochastic extinction of the fittest genotype) does not operate. In the absence of such stochastic extinction, any beneficial mutation from the fittest class would move the distribution towards higher fitness, in conflict with the steady state assumption. Hence, in infinite populations, steady state can only be achieved with $\e_c =0$. Conversely, in a finite population where the fittest class can be lost due to genetic drift, Muller's ratchet will eventually lead to a decrease in fitness if no beneficial mutations are available.  Thus we must have $\e_c>0$ to be in the dynamic steady state.

A correct description of the dynamic equilibrium state therefore requires a suitable treatment of genetic drift. These stochastic effects are particularly important in the most-fit edge of the distribution, which we call the ``nose,'' where the number of individuals with a particular fitness is small.  Our analysis is based on matching a stochastic treatment of the nose with a deterministic description of the bulk of the fitness distribution (as governed by \eq{det} with $\e >0$). The stochastic dynamics of the nose are determined by two competing processes: the random extinction of the most-fit class, versus the establishment of a new more-fit class due to a beneficial mutation. At stationarity, the rates of these two processes have to be equal; this condition determines the number of individuals in the nose class.  We match this stochastic condition with the number of individuals in the nose class calculated based on the deterministic distribution.  This determines the critical fraction of beneficial mutations $\e_c$ as a function of the population parameters.

\subsection{The shape of the fitness distribution}
\begin{figure}[tp]
\begin{center}
\includegraphics[width=0.85\columnwidth]{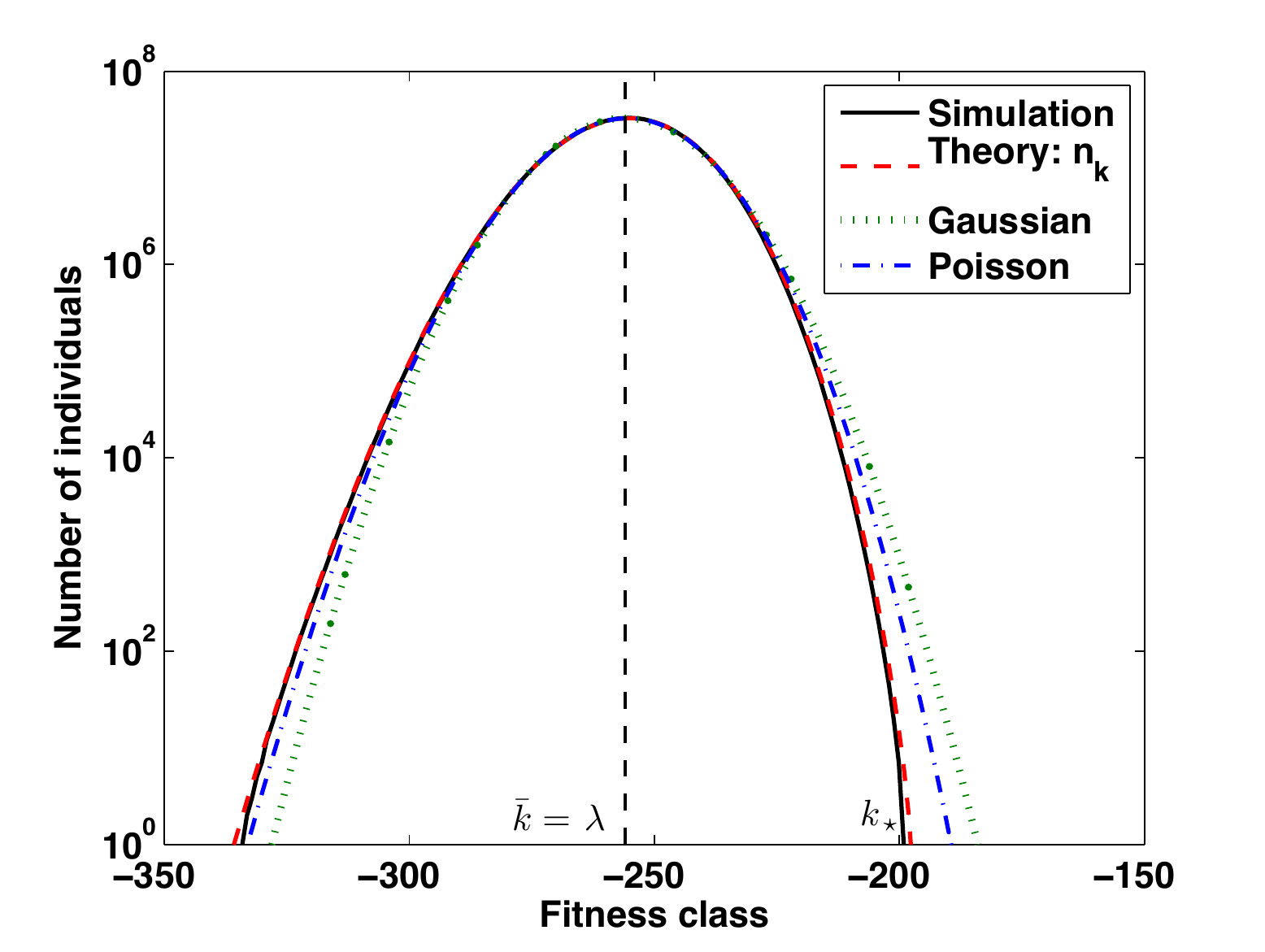}
\caption{Comparison of the fitness distribution obtained in simulations with the analytic result \eq{fast} and a Gaussian with equal mean and variance. The simulation result is the median fitness distribution observed over $10^6$ generations. \label{fig:popdis}}
\end{center}
\end{figure}

We begin with a deterministic analysis of the shape of the fitness distribution at steady state. Since only relative fitnesses matter, it is convenient to set the origin of $k$ such that the mean of the fitness distribution is $\bar k = \lambda$.  We solve for the steady state of \eq{det} using Fourier analysis (see SI).  We find \be n_k = N e^{-\lambda (1-2\e) - k \log \sqrt{{ 1-\e \over \e}} } J_{k} (2\alpha), \label{fast} \ee where $J_k$ denotes the Bessel function of order $k$ and we have defined $\alpha \equiv \lambda \sqrt{\e (1-\e)}$. A comparison of this solution to simulation results is shown in Fig.~\ref{fig:popdis}. For $k$ in the vicinity of $\bar{k}$, this distribution is approximately Gaussian with variance $\sigma_k^2 = \lambda(1-2\e)$. While the the low fitness tail of the distribution decays less rapidly than a Gaussian, the high fitness side decays more rapidly than a Gaussian.

It is important to note that in the deterministic limit there is no true stationary state for arbitrary $\e > 0$; this manifests itself in the loss of positivity of the solution given by \eq{fast} for high-fitness classes above the position of the nose class.  While the position of the nose, which we will call $k_\star$, requires a more careful treatment, \eq{fast} gives a very accurate description of the {\it bulk} of the fitness distribution (i.e. $n_k$ for $k > k_\star$).

To determine the position of the nose, $k_\star$, we observe that if classes with $k < k_\star$ carry no individuals, fitness class $k_\star$ does not receive an influx of mutations from the more-fit class $k_\star - 1$.  That is, term $III$ in \eq{det} is absent, yielding $(\bar{k} - k_\star -\lambda)n_{k_\star} + \lambda\e n_{k_\star+1} = 0$. This along with \eq{fast}, gives the following equation for $k_\star$
\be \label{kstar} k_\star J_{k_\star}(2\alpha) = \sqrt{\alpha}J_{k_\star+1}(2\alpha) \ee
%this indicates that for a given $\e$, the position of the nose is determined by the last zero of $J_{k-1}(2\alpha)$ with increasing $k$.
This must in general be determined numerically.  However, in SI we show that \be \label{x0} k_\star \approx \begin{cases} \lambda^2 \e & \textrm{for } \e \lambda^2 \ll 1  \\ 2\lambda \sqrt{\e (1-\e)} & \textrm{for } \e \lambda^2 \gg 1 . \end{cases} \ee To gain some intuition into the significance of $k_\star$, note that $k_\star=0$ in an infinite population with $\e_c=0$. In this case, the fitness distribution is the familiar Poisson distribution and the fittest class contains a fraction $e^{-\lambda}$ of all individuals. In a finite population,  $Ne^{-\lambda}$ will be smaller than one for sufficiently large $\lambda$, which implies that at steady state the mutation free genotype is typically absent and therefore $k_\star$ is larger than 0. This adjustment of $k_\star$ for finite population size is reflected in \eq{x0}, which states that the most-fit class gets closer to the population mean as $\e$ increases.

\subsection{The stochastic matching condition}
We seek to determine the $\e=\e_c$ at which the distribution is stationary for a given finite $N$. Since the most-fit class is populated by a comparatively small number of individuals, fluctuations due to genetic drift may change its occupancy significantly. In particular, the population at the nose can go extinct, in which case the next fitness class becomes the new nose. Alternatively, a lucky beneficial mutation can cause the nose to advance by one class. Stationarity can therefore only be achieved in an average sense: the rate of advancing the nose has to equal the rate of extinction of the nose.

Since these two processes depend sensitively on the number of individuals in the nose, requiring the equality of extinction and establishment rates determines the average size of the population of the nose, $n_{k_\star}$, as a function of $U$, $s$, and $\e$. We match this $n_{k_\star}$ to the deterministic solution determined above to find $\e_c(N,s,U)$.

The nature of the dynamics at the nose depends qualitatively on $n_{k_\star}$.  If $n_{k_\star}$ is large enough that its dynamics are dominated by selection, it is rarely lost and Muller's ratchet is slow. Conversely, if $n_{k_\star}$ is small, the class turns over neutrally and is easily lost and reseeded. We begin by considering the slow ratchet regime where $n_{k_\star}$ is relatively large, and then turn to the opposite fast ratchet regime.

\paragraph{Slow ratchet regime.}
In the regime where $\nz s > 1$, the fittest class is only rarely lost due to drift. Note that this regime corresponds to $\e\lam^2 < 1$, which implies $\nz\approx Ne^{-\lambda}$ and $k_\star \approx 0$.  In this regime the mean extinction rate for the nose, $r_-$, is due to rare large fluctuations which overcome the "restoring force" due to selection trying to preserve mutation selection balance. Estimates of this extinction rate have been obtained via diffusion theory \cite{Haigh78, Stephan02, Gordo00} and have the form (see SI)  $r_- \approx e^{-\gamma s \nz} \gamma s \sqrt{\gamma s\nz /\pi}$, where $\gamma$ is a parameter characterizing the effective strength of selection on the fittest class, introduced by Haigh \cite{Haigh78}.

\begin{figure}[tp]
\begin{center}
\includegraphics[width=0.99\columnwidth]{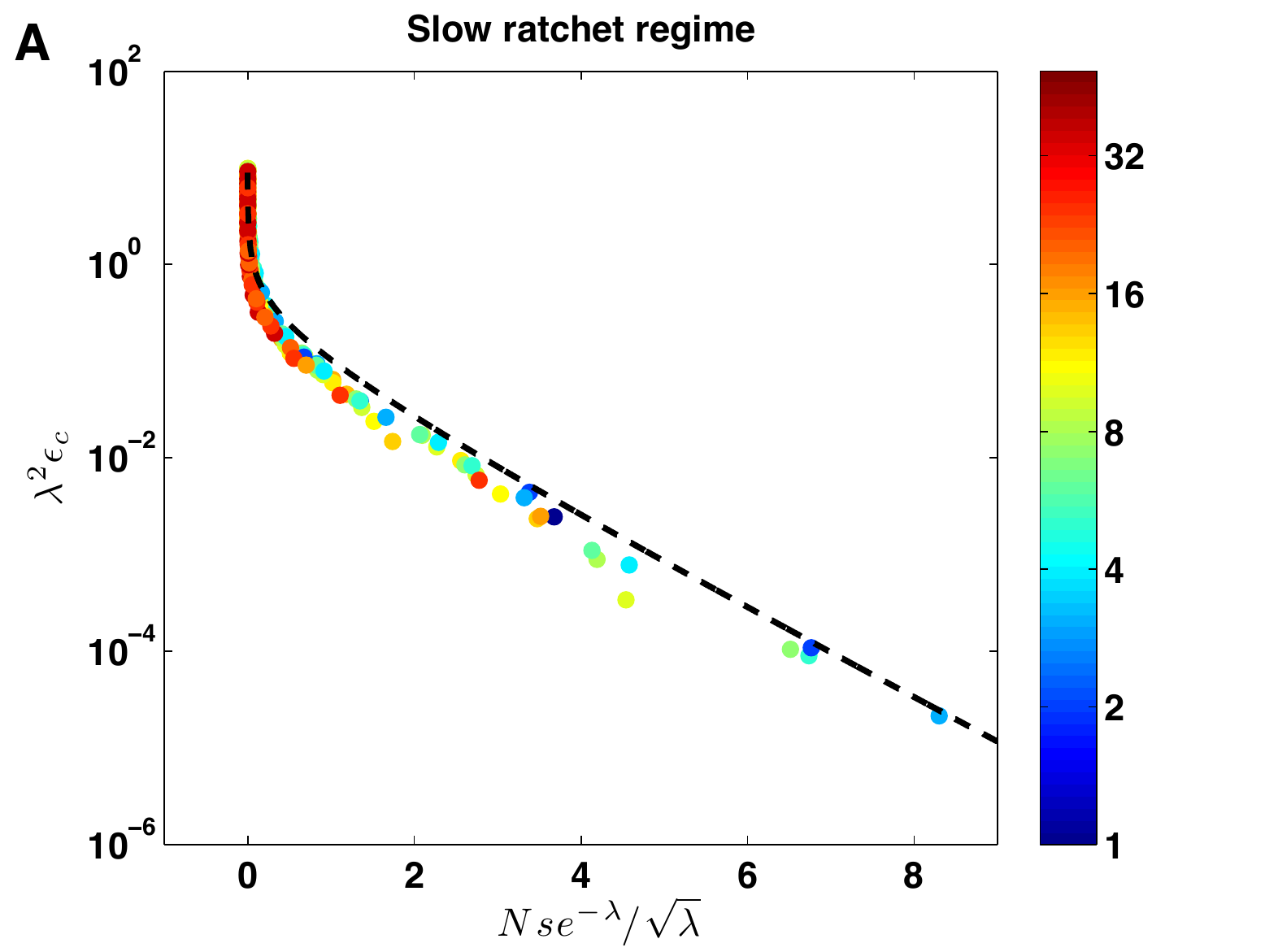}
\includegraphics[width=0.99\columnwidth]{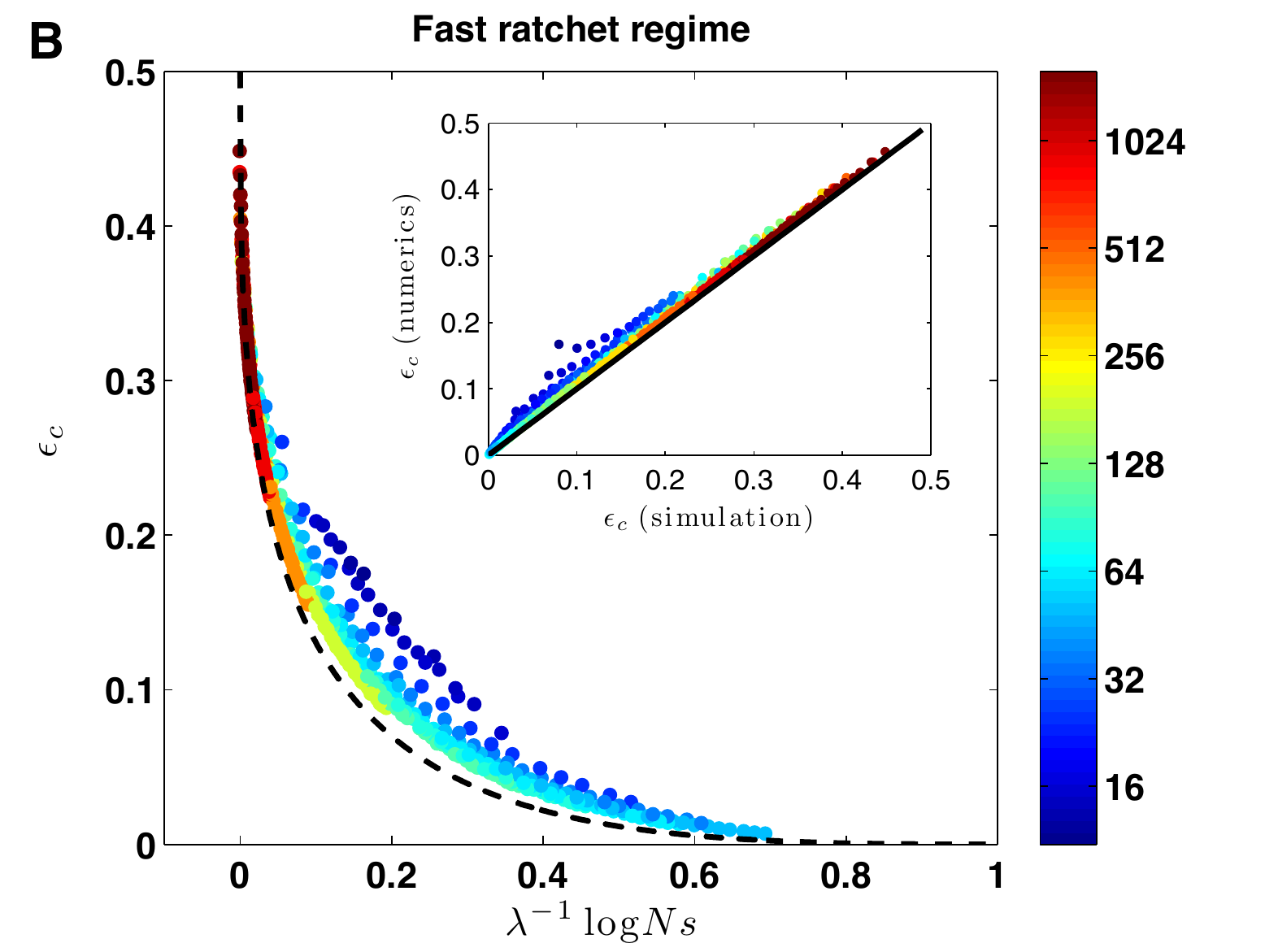}
\caption{The critical proportion of beneficial mutations, $\e_c$, across both slow and fast ratchet regimes for many combinations of $U$, $s$ and $N$ for the two regimes. Simulation results are shown as points, with color indicating $\lam = {U\over s}$. \textbf{(A)} Slow ratchet regime.  We show $\lam^2\e_c$ as a function of ${1 \over \sqrt{\lam}}sNe^{-\lam}$ for $\lam^2\e_c < 1$. The dashed black line is the analytic solution given by \eq{slow}. \textbf{(B)} Fast ratchet regime.  Note that $\e_c$ is a function of $\frac{\log Ns}{\lambda}$, and even in small populations only approaches $\e_c \approx \frac{1}{2}$ for very large $\lambda$.  The dashed black line shows the asymptotic solution $\e_c$ in the large population size limit ($\e_s\lam^2\gg1$), \eq{six}.  The inset compares the numerical solution from \eq{stoch_match} to simulation data.   \label{fig:eps_c}}
\end{center}
\end{figure}

To impose our stochastic condition, we must also calculate the rate at which a new more-fit class establishes, $r_+$. This is given by the product of the rate at which new beneficial mutants are generated and the probability they establish, $r_+ = \nz U_bP_{est}$. Since $k_\star\approx 0$ in this regime, we have $P_{est}\approx 2\D$. Equating the two rates yields the stochastic condition \be \e_c = {\gamma^2 \over 2\lambda}{e^{-\gamma s Ne^{-\lam}} \over \sqrt{\gamma s Ne^{-\lam}\pi}}. \label{slow} \ee This condition suggests that $\e_c$  in the slow ratchet regime is a function of a combination of population parameters: $\lambda \e_c(U,N,\D)/\gamma^2 \sim  f(\gamma sNe^{-\lam})$. We find that contrary to earlier studies \cite{Haigh78,  Gordo00} constant $\gamma$ is not consistent with simulations. Instead, as shown in Fig.~\ref{fig:eps_c}A the simulation data is well explained by $\gamma = 1/\sqrt{\lambda}$. Our empirical $\lambda$-dependence is however consistent with the approximation $\gamma \approx 0.6$ used by Gordo {\it et al.} \cite{Gordo00}, within the range of parameters ($0.4 \lesssim 1/\sqrt{\lambda}\lesssim 0.8$) addressed by their study.  Since we consider a much broader range of parameters ($0.1 \lesssim 1/\sqrt{\lambda} \lesssim 1$),  accounting for the dependence of $\gamma$ on $\lambda=U/s$ becomes important.  Our analytical solution \eq{slow}, shown as dotted line in Fig.~\ref{fig:eps_c}A, is in good agreement with the simulation data across this wide range of parameters.

In this regime, $\e_c$ depends exponentially on $N s e^{-\lambda}$, and hence rapidly approaches zero as $\lambda$ becomes small, approaching the infinite population limit where Muller's ratchet does not operate.

\paragraph{Fast ratchet regime.}
For larger mutation rate, smaller population size, or weaker selection, the occupancy of the most fit class decreases, thereby increasing its rate of extinction. Consequently, a higher rate of beneficial mutations (larger $\e$) is required to match the extinction rate.  The resulting rapid turnover of the population at the nose leads to the failure of the quasi-static approximation we used in the slow ratchet regime.

As the occupancy of the nose decreases, in particular when $\nz \D \lesssim 1$, the dynamics of the fittest class is governed by drift.  The rate of extinction, $r_-$, can therefore be estimated from neutral diffusion: $r_- \approx 1/ (2\nz)$. The rate at which a new more-fit class is established is given by the same formula as before, $r_+ = \e U \nz P_{est}$.  However, $P_{est}$ now refers to the probability that a new mutant lineage reaches $\frac{1}{2(k_\star+1) \D}$ individuals. At this point, the lineage crosses over from stochastic to deterministic dynamics, entering the domain described by the deterministic solution \eq{fast}. Thus $P_{est}=2(k_\star+1) \D$.  Equating $r_-=r_+$ yields the stochastic condition: \be \nz \approx \frac{1}{\sqrt{4\e U \D k_\star}} \label{stoch_match} . \ee
This condition, together with the solution \eq{fast} for the distribution, allows us to determine $\e_c(N,\ss,U)$. As shown in \fig{fig:eps_c}, our solution \eq{stoch_match} is in excellent agreement with simulations for most of the data. However, our solution overestimates $\e_c$ for populations with small population size and small $\lam$ that still satisfy $\nz \D \lesssim 1$. A better description in this regime will require a more careful analysis of fluctuations beyond the nose.

\paragraph{Large population size limit.}
Large populations can maintain a well adapted genome with $\e_c\ll \frac{1}{2}$ even for moderately large $\lambda$. In the limit of large populations and large $\lambda$, with the limit taken so that $\e_c\lambda^2 \gg 1$ while $\e_c\ll \frac{1}{2}$, the matching condition reduces to $\sqrt{\e_c} \log{\e_c} = \lambda^{-1}\log \left( Ns e^{-\lambda(1 - 2 \e_c)} \right)$. This equation has the approximate solution \eon \e_c \approx \frac{z^2}{\log^2 \left[z/\log\left(z^{-1} \right)\right]}, \qquad 2z \equiv 1 - \frac{\log Ns}{\lambda}. \label{six} \eoff  This result can be made more precise through iteration (see SI for derivations and comparisons to simulation data). Note here that $\e_c$ depends only very weakly on $Ns$ and on $\lambda$, in contrast to the slow ratchet regime where $\e_c$ declines exponentially with $Ns e^{-\lambda}$.

\paragraph{High mutation rate or weak purifying selection.}
In this limit, the dynamic equilibrium tends towards the state where beneficial and deleterious mutations are equally frequent, $\e = \frac{1}{2}$.  This corresponds to a population that can no longer maintain a well-adapted state.  In the SI, we derive an approximate expression for $\e_c$ in the fast-ratchet regime in the limit of large $\lambda$: \begin{equation} \label{eq:eps_c_x} \e_c \approx \frac{1}{2}-\left(\frac{3}{4\lambda} \log Ns\right)^{1/3} . \end{equation}  From this expression we can immediately determine the point at which the population can no longer maintain a well-adapted state, as defined by a maximum fraction of beneficial mutations, $\e_\max$.  In order to have $\e_c < \e_\max$, we require \eon \frac{\lambda}{\log Ns} \leq \frac{3}{4} \left[ \frac{1}{0.5 - \e_\max} \right]^3, \eoff valid for $\e_\max \gtrsim 0.2$.  Note that for $\e_\max = 1/4$, the RHS of this expression is $48$; for $\e_\max = 1/3$, the RHS is 162.  Thus even small populations with $Ns \gtrsim 1$ can maintain relatively well-adapted genomes in the face of high mutation rates and weak selection (i.e. when $U \gg s$).

\section{Discussion}
In infinite populations it has long been recognized that the balance between mutational pressure and purifying selection leads to a fitness equilibrium \cite{Haigh78, Eigen71}.  Our analysis demonstrates that such an equilibrium also exists in a finite population, despite the action of genetic drift and Muller's ratchet.  Previous work on these finite populations has focused primarily on the rate of Muller's ratchet in the absence of beneficial mutations, and has generally assumed that back and compensatory mutations are rare enough to be neglected \cite{Etheridge07, Pfaffelhuber11,Stephan02, Gordo00, Gordo00b, Gordo04}.  However, we expect that beneficial mutations will become more common as a population accumulates deleterious mutations.  Our analysis has illustrated how these beneficial mutations balance Muller's ratchet in asexual populations, to create a dynamic equilibrium state.

\begin{figure}[t]
\begin{center}
\includegraphics[width=0.9\columnwidth]{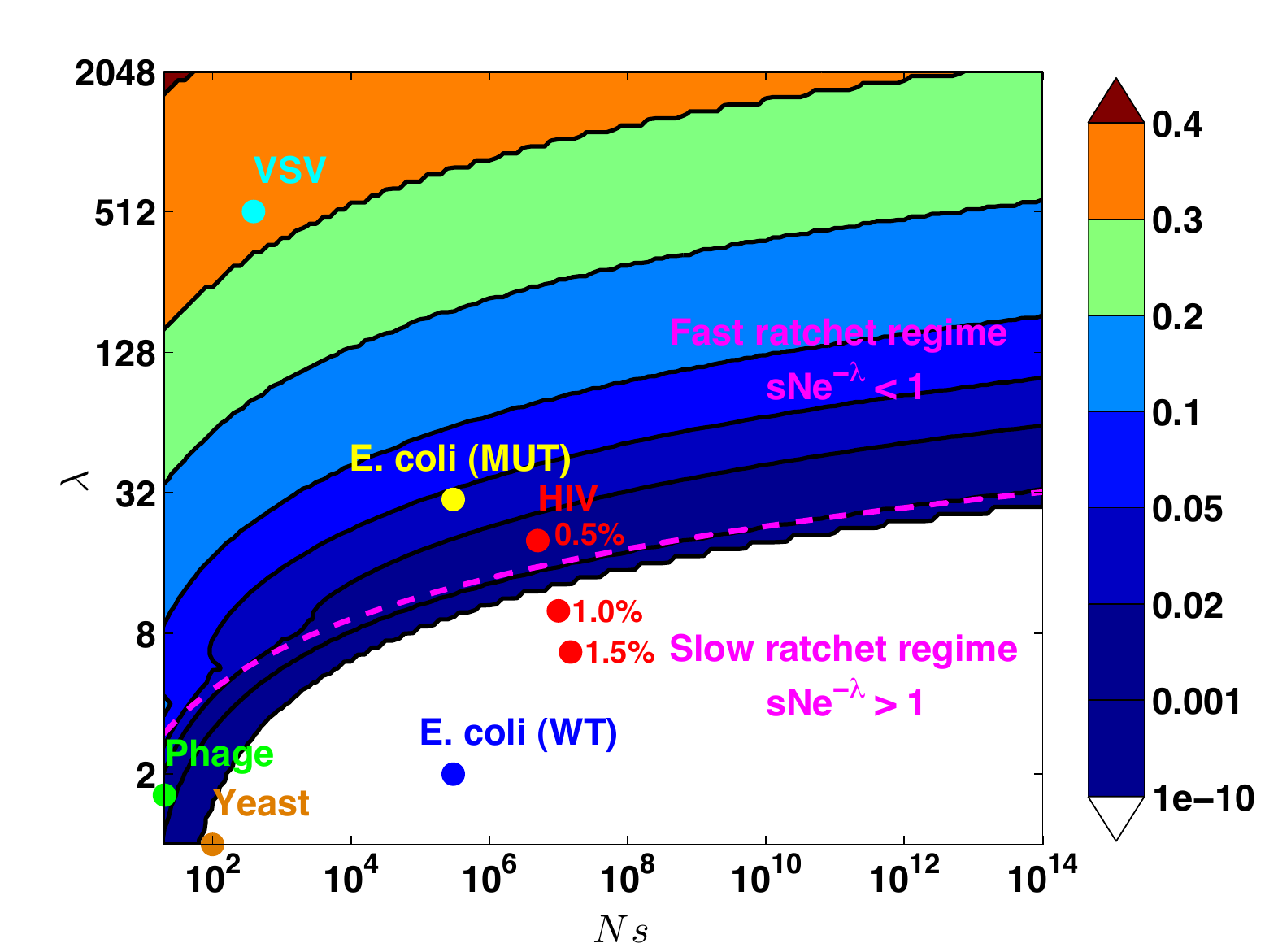}
\caption{The fraction of beneficial mutations $\e_c$ necessary to maintain the dynamic mutation selection balance in a population with parameters $Ns$ and $\lambda = {U\over s}$. The dotted line separates the slow and fast ratchet regimes. Experiments with various model organisms with different population parameters are represented as points. \label{fig:phasediagram}}
\end{center}
\end{figure}

We have shown that for any population size, mutation rate, and selection pressure, this dynamic equilibrium state is characterized by the proportion of beneficial mutations, $\e_c$, necessary to balance the ratchet. This $\e_c$ is a measure of how optimal a genome a finite population can maintain in the presence of mutations, selection, and genetic drift. Larger $\e_c$ indicates that the population needs more beneficial mutations to be in the stable state, and hence must be less fit, and vice versa. In the limit $\e_c \to 0$, the population is perfectly adapted, and no beneficial mutations are possible.  The opposite case of mutational meltdown, where a population cannot maintain an adapted genome, corresponds to $\e_c \to 1/2$, where a random mutation is as likely to be beneficial as deleterious.  There are two qualitatively different regimes for $\e_c$, as shown in \fig{fig:phasediagram}. In the slow ratchet regime, selection stabilizes the nose, so $\e_c$ is small and depends exponentially on population parameters $Ns$ and $\lam$. On the other hand, in the fast ratchet regime, $\e_c$ must be larger to balance the faster ratchet, and depends more weakly on the population parameters $Ns$ and $\lam$. The boundary of the two regimes is roughly given by $N s e^{-\lambda} = 1$.

We note that Rouzine \et.~\cite{Rouzine03,Rouzine08} studied population dynamics under the combined action of purifying and positive selection in a model similar to ours, finding both simulation and analytical evidence for the existence of the dynamic balance state.  However, their analysis focuses on the rate of the ratchet (with $\e \ll \e_c$) or the rate of adaptation (with $\e \gg \e_c$), and does not apply as $\e \to \e_c$.  Our work, by contrast, focuses exclusively on the dynamic equilibrium state and the critical $\e_c$ required to maintain it.

Our analytic results for $\e_c$ in the dynamic equilibrium were made possible by the explicit calculation of the shape of this fitness distribution, given in \eq{fast} and illustrated in \fig{fig:popdis}. With increasing $U/s$ and decreasing $N$ this distribution deviates from the Poisson distribution for the classic mutation-selection balance \cite{Haigh78}, most notably reducing the fitness of the top class, $k_\star$, relative to the mean. Examining the properties of the dynamic mutation-selection equilibrium over the full parameter range, shown in \fig{fig:phasediagram}, has revealed a strong asymmetry between beneficial and deleterious mutations. For example, for a small population $Ns\sim10^3$ with high mutation rate $U/s \sim 8$, just $2\%$ beneficial mutations are enough to counteract the effect of deleterious mutations.  This indicates that purifying selection is remarkably effective even for conditions where Muller's ratchet would proceed extremely quickly in the absence of back and compensatory mutations.

Our analysis implies that populations with $\e > \e_c$ should adapt, while populations with $\e < \e_c$ should decline in fitness.  Experimental evolution of model organisms in controlled laboratory environments appears to be consistent with this expectation.  In particular, Silander \et. \cite{Silander07} showed that bacteriophage $\phi X 174$ converged to a population-size dependent fitness plateau, as our model would predict.  For their population parameters ($s \sim 0.08$, $U \sim 0.13$) our theory predicts $\e_c$ ranging from $2\%$ to $0.2\%$ for their experiments with $N$ ranging from $100$ to $200$, consistent with the beneficial mutation rates they infer. Their experiments with lower population sizes ($N\lesssim 30$) are in the Fast ratchet regime and estimating $\e_c$ requires further analysis of fluctuations beyond the nose.

Other experiments with vesicular stomatitis virus found that large populations adapt while small populations melt, also consistent with our analysis \cite{Elena00}.  Similar results have been observed in yeast \cite{Lang11, Desai07}.  In the long-term evolution experiments of Lenski and collaborators in \emph{E. coli}, parameters are such that we expect the critical $\e_c$ to be very small, $\e_c \ll 10^{-10}$, consistent with their observation of continuous adaptation even after tens of thousands of generations.  On the other hand, mutator strains of \ecoli shows decrease in fitness \cite{Sniegowski11}, which could be used to further test the model. In \fig{fig:phasediagram}, we show rough estimates for the parameter regimes in which each of these experimental systems lie.

Population parameters for HIV place it close to the boundary between the slow and fast ratchet regimes (\fig{fig:phasediagram}), so that its evolutionary dynamics depend sensitively on the strength of selection. This intriguing case highlights two important issues.  First, our analysis has assumed that all mutations have the same selective effect $\ss$, illustrating how well a particular strength of purifying selection can maintain well-adapted states.  In general, however, mutations have a range of selective effects.  Mutations with a particular characteristic strength may tend to dominate the dynamics, if more strongly selected mutations are quickly eliminated by selection while more weakly selected mutations are effectively neutral given interference from selection on the other mutations. However, the effect is non-trivial and generalizing our results to an arbitrary distribution of mutational effects remains an important avenue for future work.  Second, our analysis has focused on asexual populations, and neglected recombination.  Since even weak recombination has the potential to significantly slow Muller's ratchet, it would be interesting to generalize our dynamic balance to include its effects.  Increasing recombination rates would presumably allow the population to maintain better-adapted genotypes in this steady state.

We have argued that the dynamic mutation selection balance is a stable evolutionary attractor. If the population size is independent of $\e$, this is immediately obvious from \fig{fig:fitlandscape}. Under some conditions, however, the population size might be a function of absolute fitness and hence of $\e$. Muller's ratchet might then lead to a decrease in fitness, resulting in a smaller population size, which in turn leads to a faster ratchet. This phenomenon of mutational meltdown might cause rapid population extinction \cite{Lynch93}. \eq{eq:eps_c_x}, however, shows that the degree of adaptation depends only weakly on the population size and a stable equilibrium can be maintained even if $N$ depends on $\e$ rather strongly.

The dynamic balance state has important implications for patterns of molecular evolution.  Recent analysis of the effects of purifying selection on the structure of genealogies has suggested that Muller's ratchet plays a crucial role in determining the structure of genetic variation in asexual populations or on short distance scales in the genomes of sexual organisms \cite{Gordo02, Seger10}. However, these expectations must be revised if populations exist instead in the dynamic state in which both beneficial and deleterious mutations fix, without any continuous net accumulation of deleterious mutations.  This means that even though no change in fitness occurs, signatures of both positive and negative selection are likely to be found in patterns of molecular evolution, as has been suggested by earlier studies \cite{Hartl96, Antezana97}.  We argue that this state is the natural null expectation for the effects of mutations and purifying selection on patterns of genetic variation; efforts to look for positive selection which represent ``true" adaptation should look for deviations from this situation.

Finally, we note that although the dynamic balance we have analyzed is stable, a population will typically fluctuate around this steady state.  A few beneficial mutations may become established at the nose by chance, leading to a temporary increase in mean fitness which is later balanced by a reduction in $\e$ at this higher fitness, restoring the population to the equilibrium state. Conversely, a few clicks of Muller's ratchet will occasionally lead to temporary reductions in population fitness before the corresponding increase in $\e$ restores the steady state.  These fluctuations around the dynamic balance may be important to the clonal structure of the population, and hence are likely to play a key role in patterns of molecular evolution and in understanding the effects of recombination.

\section{Acknowledgements}
We thank Pierre Neveu, Adel Dayarian, Aleksandra Walczak and Paul Sniegowski for many useful discussions.  SG, DB and BIS were supported by NIGMS and HFSP.  ERJ acknowledges support from an NSF graduate research fellowship.  RAN is supported by the ERC through grant StG-2010-260686.  MMD acknowledges support from the James S. McDonnell Foundation, the Alfred P. Sloan Foundation, and the Harvard Milton Fund.

\bibliographystyle{pnas}
\bibliography{mullerbackbib}

\newpage
\section{Supplemental information}
\paragraph{Simulation methods.}
Our simulations are done using a custom written Python code available on request. We implement a discrete time Wright-Fisher model where the population is represented by a vector $n_k$ with elements corresponding to the number of individuals in fitness class $k$. Each generation consists of separate selection and mutation steps. To implement selection, the vector $n_k$ is multiplied by $e^{\D(k-\bar{k})-\alpha}$ to obtain a vector $\tilde{n}_k$ containing the expected number of offspring in class $k$. Here $(k-\bar{k})$ is the fitness of class $k$ relative to the population mean, and $\alpha = {\frac{\sum{n_k}-N}{N}}$ maintains an approximately constant population size around $N$.

Our model is simplified by using beneficial and deleterious mutation with effect $\pm \D$ and mutation rate $U\e$ and $U(1-\e)$, respectively. To implement mutation, we calculate the probability $P(i,j)$ of a genome being hit by $i$ beneficial and $j$ deleterious mutations, which are Poisson distributed. This mutation matrix is then applied to $\tilde{n}_k$, i.e. the parts of $\tilde{n}_k$ are moved up or down according to the net number of mutations they accrued in this time step. Having constructed the expected number of individuals in fitness class $k$ after selection and mutation, we draw a population sample from each class from a Poisson distribution. If necessary, the population is re-centered in the discrete vector $n_k$.  This prevents occupied classes from running off the grid due to accumulated increase or decrease in their absolute fitness.

The above process is repeated for a specified number of generations. The speed of adaptation and other features of the dynamics are measured after an equilibration time to remove transient effects from the initial conditions. In the parameter regimes studied, we found that $10^4$ generations was generally sufficient.

To solve for $\e_c(U,N, \D)$, we re-run the simulation while iteratively adjusting $\e$ to get $v$ as close to zero as possible.  Since $v$ is a stochastic quantity,  $\e_c$ can only be determined with limited accuracy, but this can be improved by increasing the number of generations for each run.  We have also run simulations in which $\e$ varies with fitness (i.e. with $k$) to demonstrate that the population indeed evolves towards the values of $k$ at which they have the appropriate $\e_c$.

\paragraph{Stationary distribution in the presence of deleterious and beneficial mutations.}
The steady state distribution satisfies
\be
0 = (k - \bar{k})n_k-\lambda n_k + \lambda(1-\e)n_{k+1} + \lambda\e
n_{k-1}   \label{app1}
\ee
which we solve by introducing a Fourier transform
\begin{equation}
\hat{n} (q)=\sum_k e^{-iqk} n_k
\end{equation}
which yields an ODE in $\hat{n}(q)$
\begin{equation}
{d \over dq} \hat{n} (q) = -i {\lambda} [(1-\e)e^{-iq} +\e e^{iq} ] \hat{n} (q).
\end{equation}
The ODE is readily solved to give
\begin{equation}
\ln  \hat{n} (q) = {\lambda} \left [ (1-\e)  e^{-iq} -  \e e^{iq} \right ]-{\lambda}(1-2\e)
\end{equation}
where the last term is a constant added to enforce normalization $\sum_k n_k =1$.

Fourier transforming back to solve for $n_k$ gives
\begin{equation}
n_k= e^{-{\lambda} (1-2\e)} \int_{-\pi}^{\pi} {dq \over 2\pi} \exp \left [ ikq + {\lambda} (1-\e)  e^{-iq} -{\lambda} \e   e^{iq} \right ] .
\end{equation}

This can be re-written by shifting the integration contour upward into the complex plane by $Q_*=\ln \sqrt{(1-\e)/\e}$, i.e. change integration variable $q=q'-iQ_*$ which yields
\begin{eqnarray}
n_{k} & = & e^{-{\lambda}(1-2\e)-k \log \sqrt{{ 1-\e \over \e}}}
\int_{-\pi}^{\pi} {dq' \over 2\pi} e^{  \left \{   -ikq' +  \alpha [ e^{-iq'} -    e^{iq'} ] \right \} } \\ \nonumber
&=& e^{-{\lambda}(1-2\e)-k \log \sqrt{{ 1-\e \over \e}}}
J_k(2\alpha)
\end{eqnarray}
where $\alpha\equiv{\lambda} \sqrt{\e (1-\e) }$ and the Bessel function is identified from its integral representation \cite{Abramowitz}.

While this solution holds without any restriction on parameters, it is instructive to examine the $\e \rightarrow 0$ limit, for which we have a regular perturbative expansion in $\e {\lambda}^2 <<1$:
 \begin{eqnarray}
 &n_{k}=e^{-{\lambda}(1-2\e)-k \log [{\lambda}( 1-\e )]}
 \sum_{j=0}^{\infty} { (-1)^j\alpha^{2j} \over j!(k+j)!} \\
&= {[{\lambda}( 1-\e )]^k e^{-{\lambda}(1-2\e)} \over k!}
 \left \{ 1 - {{\lambda}^2 \e (1-\e) \over k+1 } +...\right \}
 \end{eqnarray}
 which in the $\e \rightarrow 0$ recovers the familiar Poisson mutation/selection balance.

\paragraph{Rate of extinction of the nose in slow ratchet regime.}
The rate of extinction for the nose can be calculated using the stochastic dynamics of
the fittest class, which can approximately be modeled by a diffusion equation:
\be
\partial_t p(x,t) =  -\partial_x\left[D_1(x)p(x,t)\right] + \partial_x^2\left[D_2(x)p(x,t)\right]
\ee
where $p(x) = p(n_{k_\star}/N=x)$ is the probability distribution of the nose occupancy and $x_{k_\star} = e^{-U/s}$ is the mean fraction of the population, $D_1(x) = \hs x(1-x/k_\star)$, and $D_2(x) = x(1-x)/2N$. Both $x=1$ (fixation) and $x=0$ (extinction) are absorbing boundary conditions \cite{Stephan02}. The density function for fixation/extinction time $\phi(t;x)$
satisfies the backward Kolmogorov equation
\be
\p_t\phi(t;x) = D_1(x)\p_x\phi(t;x) + D_2(x)\p^2_x\phi(t;x)
\ee
where $t$ is the time interval between initial state $x$ and fixation or extinction. The mean time to fixation/extinction starting at $x = y$ at $t=0$ is given by $\t(y) = \int_0^{\infty} t\phi(t;y) dt$, which satisfies
\be
-1 = D_1(y)\p_y\t(y) + D_2(y)\p^2_y\t(y) . 
\ee
Using the integrating factor $\phi(x) = e^{2\int_0^xdz{D_1(z) \over D_2(z)}}$, yields for the mean time to extinction of the fittest class:
\be
\t(x_{k_\star}) = \int_{0}^{x_{k_\star}}dy \frac{1}{\phi(y)}\int_y^1d\zeta\frac{\phi(\zeta)}{D_2(\zeta)} . 
\ee
The second integral in the limit $Ns\gg 1$ can be approximated as
\be
2N\int_y^1d\zeta
\frac{(1-\zeta)^{\frac{2Ns(1-x_{k_\star})}{x_{k_\star}}}e^{\frac{2Ns\zeta}{x_{k_\star}}}}{\zeta(1-\zeta)}
\approx 2N\int_y^1 d\zeta\frac{e^{2Nsx_{k_\star}\left(\frac{2\zeta}{x_{k_\star}}
-\frac{\zeta^2}{x_{k_\star}^2}\right)}}{\zeta} . 
\ee
Using $\alpha = Nsx_{k_\star}, n_{k_\star} = Nx_{k_\star}$ and changing the variables,  $\phi = y/x_{k_\star} -1, z = \zeta/x_{k_\star} -1$ yields:
\be
\t(x_{k_star}) =
2n_{k_\star}\int_{-1}^0 d\phi e^{\alpha\phi^2}\int_{\phi}^{1/k_\star-1}dz \frac{e^{-\alpha
z^2}}{1+z} . 
\ee
For $\alpha\gg 1$ the integral the second integral can be readily approximated yielding
\be
\t(x_{k_\star}) \approx
2n_{k_\star}\sqrt{\frac{\pi}{4\alpha}}\int_{-1}^0d\phi e^{\alpha\phi^2}\left[erf(\sqrt{\alpha}\beta)
-erf(\sqrt{\alpha} \phi) \right]
\ee
where $\beta = {1\over k_\star} -1$. Using $\phi^2 = 1 - \theta^2/\alpha$ allows the evaluation of the integral in the $\alpha\gg 1$ limit and yields
\be
\t(x_{k_\star}) \approx n_{k_\star}\sqrt\pi\alpha^{-3/2} e^{\alpha} . 
\ee

Assuming an exponential distribution for time of extinction of the fittest class, the rate of extinction is given by $r_- = 1/\t(x_{k_\star})$, which is used in the Main Text.

\begin{figure}
\begin{center}
  \includegraphics[width=0.9\columnwidth]{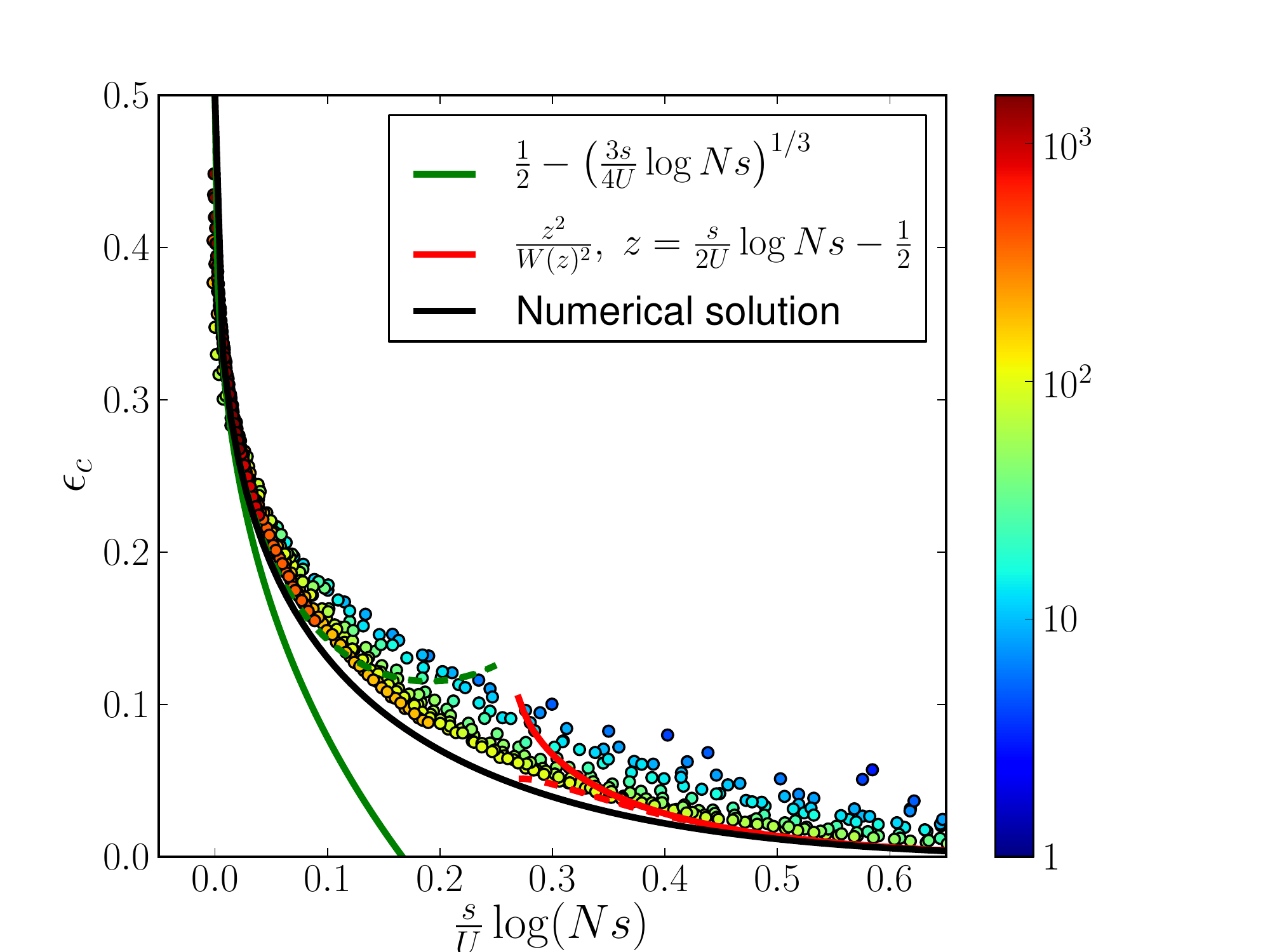}
  \caption{Dynamic mutation-selection balance. The figure shows $\e_c$ for many
combinations of $U$, $s$ and $N$ as a function of $\frac{s}{U}\log Ns$, while
the color codes for $\e_c U^2 s^{-2}$. If $\e_c U^2 s^{-2}\gg 1$,
$\e_c$ is solely a function of $\frac{s}{U}\log Ns$ and is well described
by the numerical solution of Eq. \ref{stoch_match} shown as a black line. The
asymptotic approximations for large $\e_c$ (Eq. \ref{eq:large_eps}) and small
$\e_c$ (Eq. \ref{eq:small_eps}) are shown as green and red lines,
respectively. The dashed lines correspond to the more accurate version mentioned
in the main text.
\label{fig:eps_asymptotic}}
\end{center}
\end{figure}

\paragraph{Asymptotic expressions for $\e_c$}
The matching condition used to determine $\e_c$ in the fast-ratchet regime can be rewritten as follows.
\begin{equation}
\frac{1}{\sigma\sqrt{\e k_\star}} = N e^{-\lambda(1-2\lambda)+\frac{k_\star}{2} \log\frac{1-\e}{\e}}J_{k_\star}(2\alpha) . 
\end{equation}
In the limit of $\alpha\gg 1$, the zero of the Bessel function is approximately at $k_\star+1\approx \alpha$ and the Bessel function at $k_\star$ evaluates to roughly $\sim k_\star^{2/3}$ (Abramowitz 9.3.33 and 9.1.27).
Hence we have
\begin{equation}
\lambda(1-2\e)-\lambda \sqrt{\e(1-\e)}\log\frac{1-\e}{\e} = \log N s
\end{equation}
where we have neglected powers to the $\frac{1}{6}$ and  $\mathcal{O}(1)$ factors inside the logarithm.
In the limit $\e\ll 1$ (but $\lambda^2 \e\gg 1$), the matching simplifies to
\begin{equation}
\label{eq:eps}
1-2\e+\sqrt{\e} \log\e = \lambda^{-1}\log N s
\end{equation}
which simplifies further to
\begin{equation}
\sqrt{\e} \log\e \approx  \lambda^{-1}\log(Ns)-1 = -2z
\end{equation}
where we defined $z$ for convenience. This can be solved for $\e$:
\begin{equation}
\label{eq:small_eps}
\e_c = \frac{z^2}{W(-z)^2} \approx \frac{z^2}{(\log(z)-\log(-\log(z)))^2}
\end{equation}
where $W(x)$ is the $-1$ branch of Lambert's W-function, i.e.~the solutions of $W(x)e^{W(x)}=x$. The linear correction in Eq.~\ref{eq:eps} can be incorporated iteratively.
\begin{equation}
\e_{i+1} = \frac{(z+\e_i)^2}{W(-z+\e_i)^2} . 
\end{equation}
This iteration converges for small $C$ and $\e$. At larger $C$, the branch of the $W(x)$ function is lost. The result $\e_2$ obtained after the first iteration, starting with Eq.~\ref{eq:small_eps}, is also shown in Fig.~\ref{fig:eps_asymptotic} as dashed red line.

The other limit that is amenable to analytic calculations is the limit $\e \to 1/2$. To this end, we define $\delta = \frac{1}{2}-\e$ and expand the right hand side of Eq.~\ref{eq:eps}
\begin{equation}
1-2\e-\sqrt{\e(1-\e)} \log\frac{1-\e}{\e} = \frac{4}{3}\delta^3 + \frac{44}{15}\delta^5 + \mathcal{O}(\delta^7) =  \lambda^{-1}\log N s . 
\end{equation}
From this, we find
\begin{equation}
\label{eq:large_eps}
\e \approx \frac{1}{2} - \left(\frac{3}{4\lambda}\log(Ns)\right)^{1/3} . 
\end{equation}
This expression is compared to simulation results in Fig.~\ref{fig:eps_asymptotic}. The term proportional to $\delta^5$ can again be included by iteration and the result is shown as the dashed green  line in Fig.~\ref{fig:eps_asymptotic}.

\end{article}

\end{document}